# No evidence for an association between gender equality and pathogen prevalence – a comment on Varnum and Grossmann (2017)

Alexander Koplenig[1] and Sascha Wolfer[1]

1 Leibniz-Institute for the German Language (IDS), Mannheim, Germany.

Correspondence: koplenig@ids-mannheim.de

In a previous study published in *Nature Human Behaviour*[1], Varnum and Grossmann claim that reductions in gender inequality are linked to reductions in pathogen prevalence in the United States between 1951 and 2013. Since the statistical methods used by Varnum and Grossmann are known to induce (seemingly) significant correlations between unrelated time series[2–6], so-called spurious or non-sense correlations, we test here whether the statistical association between gender inequality and pathogens prevalence in its current form also is the result of mis-specified models that do not correctly account for the temporal structure of the data. Our analysis clearly suggests that this is the case. We then discuss and apply several standard approaches of modelling time-series processes in the data and show that there is, at least as of now, no support for a statistical association between gender inequality and pathogen prevalence.

To test the role of pathogen prevalence, Varnum and Grossmann report the standard Pearson correlation coefficient $r_{xy}$ that can be calculated based on a simple linear regression model:

$$y_t = \beta_0 + \beta_1 x_t + \varepsilon_t \qquad (1)$$

where $t = 1, 2, \ldots T$ indexes years, $y_t$ represents the gender inequality index, $x_t$ represents the predictor pathogen prevalence and $\varepsilon_t$ is an independent and identically distributed error term with mean 0 and constant variance, i.e. $\varepsilon_t \sim N(0, \sigma^2)$. Based on eq. (1), $r_{xy}$ can be calculated as $\beta_1 \sigma_x / \sigma_y$ where $\sigma_x$ and $\sigma_y$ are the standard deviations of *x* and *y*. In addition, Varnum and Grossmann fit a multiple regression model that can be expressed as

$$y_t = \beta_0 + \beta_1 x_t + \beta_2 u_t + \beta_2 v_t + \beta_3 w_t + \varepsilon_t \qquad (2)$$

where, in addition to the above, $u_t$ represents climatic stress, $v_t$ represents unemployment and $w_t$ represents a binary variable that indicates whether the United States were engaged in an armed conflict in *t*.

**Table 1 | Association between gender inequality and different predictors.** In each model, the outcome, *y*, is gender inequality. Column 1: predictor *x*; column 2: Pearson correlation between *y* and *x*. column 3: *P*-value of $r_{xy}$; column 4: Durbin–Watson statistic based on eq. (1); column 5: partial correlation coefficient based on eq. (2); column 6: *P*-value of $r'_{xy}$; column 7: Durbin–Watson statistic based on eq. (2).

| Predictor | $r_{xy}$ | P | $d_{r_{xy}}$ | $r'_{xy}$ | P | $d_{r'_{xy}}$ |
|---|---|---|---|---|---|---|
| US pathogen prevalence | 0.77 | < 0.001 | 0.24 | 0.69 | < 0.001 | 0.32 |
| UK price for lighting | 0.64 | < 0.001 | 0.10 | 0.68 | < 0.001 | 0.36 |
| Sweden's miliary expenditures | 0.82 | < 0.001 | 0.23 | 0.71 | < 0.001 | 0.30 |
| Europe's share of the world population | 0.90 | < 0.001 | 0.17 | 0.78 | < 0.001 | 0.26 |
| Working hours per worker in Switzerland | 0.92 | < 0.001 | 0.23 | 0.80 | < 0.001 | 0.32 |
| Fatal airliner accidents | 0.93 | < 0.001 | 1.36 | 0.67 | < 0.001 | 0.61 |
| Share of people living in rural areas in China | 0.95 | < 0.001 | 0.29 | 0.82 | < 0.001 | 1.03 |
| Decline in monitored wildlife populations | 0.99 | < 0.001 | 1.52 | 0.83 | < 0.001 | 1.36 |

Row 1 of Table 1 replicates the results of Varnum and Grossmann. As can be seen, there is a seemingly strong association between gender inequality and pathogen prevalence based on both eq. (1) and eq. (2).

To see why we this is a spurious relationship, we replicated the analyses with different predictors in row 2 – 8 of Table 1. Highly significant associations with gender inequality are observed for a wide variety of variables: the price for lighting in the UK, Sweden's military expenditures, Europe's share of the world population, the number of working hours per worker in Switzerland, the number of fatal airliner accidents, the share of people living in rural areas in China or the estimated decline in monitored wildlife populations. Given that in most of those examples, the strength of the computed associations are stronger than the strength of the relationship reported by Varnum and Grossmann, we believe that it would be hard for a person without knowledge of Varnum and Grossmann's study to identify which relationship was actually reported in a paper. To *resolve* these apparent paradoxes, we report the Durbin-Watson statistic in column 4 and column 7 of Table 1 that can be calculated as:

$$d = \sum_{t=1}^{T-1}(\varepsilon_{t+1} - \varepsilon_t)^2 / \sum_{t=1}^{T} \varepsilon_t^2 \qquad (3)$$

Since *d* is very low for each model (theoretical expectation is equal to 2), this indicates that the models are severely mis-specified regardless of the strength of the observed relationship or goodness-of-fit[3]. The reason for the mis-specification is that both gender inequality and all predictors including pathogen prevalence show clear trending behaviour, i.e. the series evolve with time in one direction. Or put differently, the significant associations between the time series are only the result of the underlying trends and do not signal the existence of any relationship between them[4]. To prove that we are not just cherry-picking, we generated the following three sets of time series (details and visualizations for each set of series can be found at https://osf.io/jga4f/?view_only=ef5d36b0716c47fc814703e6bcaeedc3):

(i) 10,000 coin flips where the number of heads minus the number of tails up until each moment in time *t* serves as the value of the series.

(ii) 10,000 random walks that were generated from the following data-generating process[5]: $r_t = r_{t-1} + \varepsilon_t$ where $\varepsilon_t \sim N(0,1)$.

(iii) 32,075 time series that we extracted from the Our World in Data online platform on a wide variety of topics such as economic development, living conditions, technology adoption, agricultural production or environmental change with available information for hundreds of different geographical regions, countries, socioeconomic factors and topic-dependent categories (e.g. different types of natural disasters).

For (i) and (ii) there cannot be any meaningful connection between gender inequality and the series, since the series are random per definition. For (iii), meaningful relationships with US gender inequality seem in almost all cases, in our view, *a priori* very implausible. As gender inequality[1], all series show strong trends over time (median absolute Spearman correlation with time $|\rho|_{med} = .85$ for (iii) and $|\rho|_{med} = .63$ for (i) and (ii)). We then calculated $r_{xy}$ between gender inequality and each of the 52,075 series. Plots (a – b) of Fig. 1 show that spurious correlations are the rule rather than the exception[3] for the random time series as significant correlations (at $P < 0.01$) occur in more than 77% of all cases for both sets. Plot (c) of Fig. 3 complements this result by showing that significant associations with gender inequality are observed in more than 82% of all real-world time series. While it might be possible "to rationalize very nearly everything"[2], the full table of all correlations (available at https://osf.io/jga4f/?view_only=ef5d36b0716c47fc814703e6bcaeedc3) clearly shows that the overwhelming majority of correlations with gender inequality are undeniably spurious, e.g. a median absolute Pearson correlation of 0.91 for 217 series that quantify the yearly meat production in different countries and geographical regions (95.85% are significant).

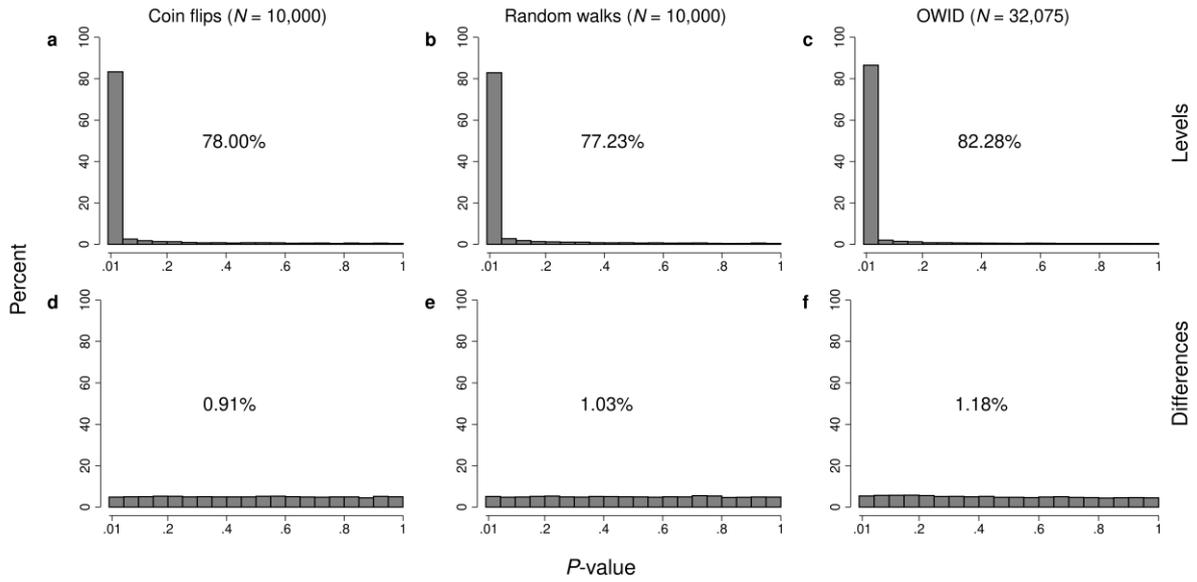

**Figure 1 | Histograms visualizing the distribution of *P*-values for the correlations between gender equality and each time series of each of the four sets.** The value in the centre of each plot represents the percentage of cases where *P* < 0.01. Plots (**a** – **c**) show the distributions for levels, plots (**d** – **f**) show the distributions for changes.

While there are more sophisticated models for time-series data[7], a simple recommendation[3,5,6] in such cases is to fit the model in changes instead of levels by taking first differences of the variable involved in the equation:

$$\Delta y_t = \beta_0 + \beta_1 \Delta x_t + \varepsilon_t \qquad (4)$$

where $\Delta y_t = y_t - y_{t-1}$; likewise for $x_t$. Plots (d – f) of Fig. 1 show that this approach solves the problem of spuriousness, in line with theoretical expectations, significant results (at the nominal 1% level) only occur in 1.07% of all cases.

If we fit eq. (4) for the association between gender inequality and pathogen prevalence, the relationship points in the opposite direction as argued by Varnum and Grossmann with $r_{xy} = -0.22$, but does not reach statistical significance ($P = 0.09$). Another recommendation to alleviate the problem of spuriousness is to include a lagged dependent variable into the equation[5], i.e. eq. (2) can be modified as:

$$y_t = \beta_0 + \beta_1 x_t + \beta_2 u_t + \beta_2 v_t + \beta_3 w_t + \beta_4 y_{t-1} + \varepsilon_t \qquad (5)$$

Fitting (5) for the association between gender inequality and pathogen prevalence yields a partial correlation coefficient that is effectively zero ($r'_{xy} = 0.01$, $P = 0.64$).

It is worth pointing out that the problem of spuriousness also affects the reliability of the gender-inequality index constructed by Varnum and Grossmann: while the correlation between levels of the number of women in the US Congress and levels of US wage inequality based on eq. (1) is $r_{xy} = 0.94$ ($P < 0.001$), the corresponding correlation is $r_{xy} = 0.14$ ($P = 0.27$) based on eq. (4) and $r'_{xy} = 0.02$ ($P = 0.29$) based on a model that includes a lagged dependent variable.

Finally, we used a model selection algorithm[8] to find the best-fitting parameters $p$, $d$, $q$ of a dynamic regression model with ARIMA($p$, $d$, $q$) errors[7], where gender inequality is the outcome and pathogen prevalence the predictor. The algorithm selects an ARIMA(0 1 1) that can be written as:

$$\Delta y_t = \beta_0 + \beta_1 \Delta x_t + \theta \varepsilon_{t-1} + \varepsilon_t \qquad (6)$$

where, in addition to the above, $\theta$ is the first-order moving-average parameter. Here, the coefficient of pathogens is marginally significant ($P = 0.07$), but negative ($\beta_1 = -0.0008$). While this would suggest a relationship in the opposite direction as argued by Varnum and Grossmann – as fewer pathogens compared to last year would be indicative of an increase in gender inequality compared to last year – we believe that such models offer a valuable method to test a potential relationship between gender inequality, pathogen prevalence and other potential predictors. We therefore encourage Varnum and Grossmann, who suggested the model selection algorithm to us, to further pursue this idea.

## Data availability

The original Varnum and Grossmann dataset is available at https://osf.io/s3pft. All time series that were taken from https://ourworldindata.org are described and cited at https://osf.io/jga4f/?view_only=ef5d36b0716c47fc814703e6bcaeedc3. Data on women in the US congress are taken from https://digital.library.unt.edu/ark:/67531/metadc267886/m1/1/high_res_d/RL30261_2013Sep26.pdf. Data on wage inequality are taken from https://www.infoplease.com/us/society-culture/gender-sexuality/womens-earnings-percentage-mens-1951-2013. The idea of generating random coin flips is mentioned in a commentary at https://stats.stackexchange.com/questions/133155/how-to-use-pearson-correlation-correctly-with-time series.

## Code availability

The code necessary to conduct the reported analysis is available in two different programming languages (Stata 14.2[9] and R[10]) on the Open Science Framework (https://osf.io/jga4f/?view_only=ef5d36b0716c47fc814703e6bcaeedc3).

## Author contributions

AK and SW designed and conceptualized the study and analysed the data. AK wrote the paper.

## Competing interests

The authors declare no competing interests.

## Acknowledgments

We thank both Michael E.W. Varnum and Igor Grossmann for responding to our request for clarification and for an open and friendly discussion. Both authors actively supported us in writing this commentary and explicitly expressed their appreciation for our focus on the discussed issues. We can only applaud them for their commitment.